**Title**

# First principles study of band line up at defective metal-oxide interface: oxygen point defects at Al/$SiO_2$ interface

**Authors**

Eric Tea[1], Jianqiu Huang[1] and Celine Hin[1,2]

[1] Department of Mechanical Engineering, Virginia Tech, Goodwin Hall, 635 Prices Fork Road - MC 0238, Blacksburg, VA 24061, USA

[2] Department of Material Science and Engineering, Virginia Tech, Goodwin Hall, 635 Prices Fork Road - MC 0238, Blacksburg, VA 24061, USA

**Abstract**

The dielectric breakdown at metal-oxide interfaces is a critical electronic device failure mechanism. Electronic tunneling through dielectric layers is a well-accepted explanation for this phenomenon. Theoretical band alignment studies, providing information about tunneling, have already been conducted in the literature for metal-oxide interfaces. However, most of the time materials were assumed defect free. Oxygen vacancies being very common in oxides, their effect on band lineup is of prime importance in understanding electron tunneling in realistic materials and devices. This work explores the effect of oxygen vacancy and oxygen di-vacancy at the Al/$SiO_2$ interface on the band line up within Density Functional Theory using PBE0 hybrid exchange and correlation functional. It is found that the presence of defects at the interface, and their charge state, strongly alters the band line up.

## 1) Introduction

The dielectric breakdown is an old problem that has been studied over decades but still remains open. Dielectric breakdown causes severe and irreversible damage to electronic devices such as capacitors or Metal-Oxide-Semiconductor devices, giving scientists a strong incentive to understand its origin. Different factors can contribute to breakdown such as temperature, pressure and humidity, or defects. [1,2] When the applied voltage on a device exceeds a limit, called the breakdown voltage, the resistance of an insulator will rapidly decrease so that it becomes temporary or permanently conductive. [3] One familiar example of this phenomenon is lightning during thunderstorms.

Previous studies showed that electronic conduction through the dielectric, which impedes the device performance, can take place via electron tunneling through the oxide barrier. It was also shown that excessive tunneling is a possible explanation for the breakdown. [32-35] The transmission coefficient $T$ for an electron of energy $E$ tunneling through a rectangular barrier of energy $U$ is given by [4]

$$T \cong e^{-2d\sqrt{\frac{2m(U-E)}{\hbar^2}}} \qquad (1)$$

where $d$ is the barrier thickness, $m$ the electron effective mass, and $\hbar$ the reduced Plank constant. The potential barrier height encountered by the incoming electron is $\Phi = U - E$. Current technology trends aim at the shrinkage of device size, going from the micro-scale to the nano-scale. Dielectric layers used in nowadays devices are therefore very thin, only a few nanometers thick [36]. In this context, the potential barrier height $\Phi$ is the only degree of freedom left that can control the tunneling current. Thus, studying band alignment at metal-oxide interfaces can provide critical information on potential barrier heights and tunnel currents in technologically important materials. [5,6]

$SiO_2$ is one of the most widely used insulators in the semiconductor device industry. Under its $\alpha$-quartz structure, it exhibits a very large band gap of 8.9 eV. [7,8] Aluminum is one of the most used metal to form electric contacts in devices. Therefore, $Al/SiO_2$ metal oxide interface has been chosen for this study since it is the most representative interface for electronic device applications.

It has been shown that oxygen vacancies are the most encountered point defect in $SiO_2$ grown onto Si substrate. [27-28] Moreover, oxygen vacancies can act as electron-traps and form conduction paths leading to leaking current. [9,10,17,22] Therefore, we focused on oxygen vacancies and how they alter band line up at Al/$SiO_2$ metal oxide interfaces. Oxygen vacancy ($V_o$) and oxygen di-vacancy ($V_{oo}$) have been studied in the following.

**2) Methodology**

All calculations have been performed in the framework of first principles Density Functional Theory. The Projector Augmented Wave method [25] has been used as implemented in the Vienna *ab initio* Simulation Package. [23,24] The Local Density Approximation (LDA) has been used for the exchange and correlation functional. [26] However, since underestimation of band gaps is a known deficiency of this approximation, [5,11] LDA results have been tested against PBE0 hybrid functional results. The plane wave cutoff energy for our calculations was set to 400 eV, and total energy convergence criteria to $10^{-5}$ eV. The Quasi-Newton algorithm has been used to relax the cells and atomic positions until forces on atoms are smaller than 1 meV/Å. This computational setup provides a very reasonable convergence of the structures and total energies. Atomic structures and charges densities are visualized with VESTA [21].

Al/$SiO_2$ metal oxide interfaces connecting Al (111) to $SiO_2$ (001) have been studied using supercell calculations. The resulting lattice mismatch at the interface is less than 1% so that stress effects can be neglected. Two types of supercell have been investigated. The first supercell type, later referred to as superlattice model, contains 7 Al layers and 10 $SiO_2$ layers. Due to periodic boundary conditions, it contains two equivalent metal/oxide interfaces and is sketched on Figure 1a. The second supercell type, later referred to as the slab model, also contains 7 Al layers and 10 $SiO_2$ layers. However, as shown on Figure 1b, it contains a ~15 Å wide vacuum region separating periodic images, and therefore only contains one interface. $SiO_2$ and Al surface dangling bonds have been passivated in order to lock charges in their bulk-like configuration away from the metal-oxide interface. Hydrogen atoms and 0.75 fractional hydrogen atoms have been used to passivate $SiO_2$ and Al surface dangling bonds respectively. The amount of charge needed for passivation has been deducted from electron and bonds counting.

In defective interface calculations, oxygen atoms have been removed from the supercells. For oxygen vacancies, one oxygen atom was removed from $SiO_2$ at the interface. For oxygen di-vacancies, two oxygen atoms bonded to the same Si atom were removed at the interface. Both vacancy types have been studied at different charge states (±3, ±2 and ±1). Defect formation energies have been calculated as

$$E_f = E_{defect} - E_{perfect} + nE_{atom} + q(\mu_e + E_{VBM}) \quad (2)$$

where $E_{defect}$ is the defective system total energy, $E_{perfect}$ is the defect-free system total energy, $n$ is the number of oxygen atoms removed from the system, $E_{atom}$ is the chemical potential of the oxygen atom, $q$ is the charge state of the system, $\mu_e$ is the electron chemical potential, and $E_{VBM}$ is the valence band maximum energy level. [14,15,16]

The interfacial defect areal concentration in our calculations is $1.2\times10^{14}cm^{-2}$, set by the supercells dimensions sketched in Figure 1. Such oxygen vacancy concentrations have been evidenced by experiments in annealed α-quartz where concentrations as large as $2.6\times10^{14}cm^{-2}$ have been measured [40]. Annealing is performed when preparing open surfaces for subsequent material growth. Therefore, we believe that our oxygen vacancy concentration is representative of a realistic defective interface material.

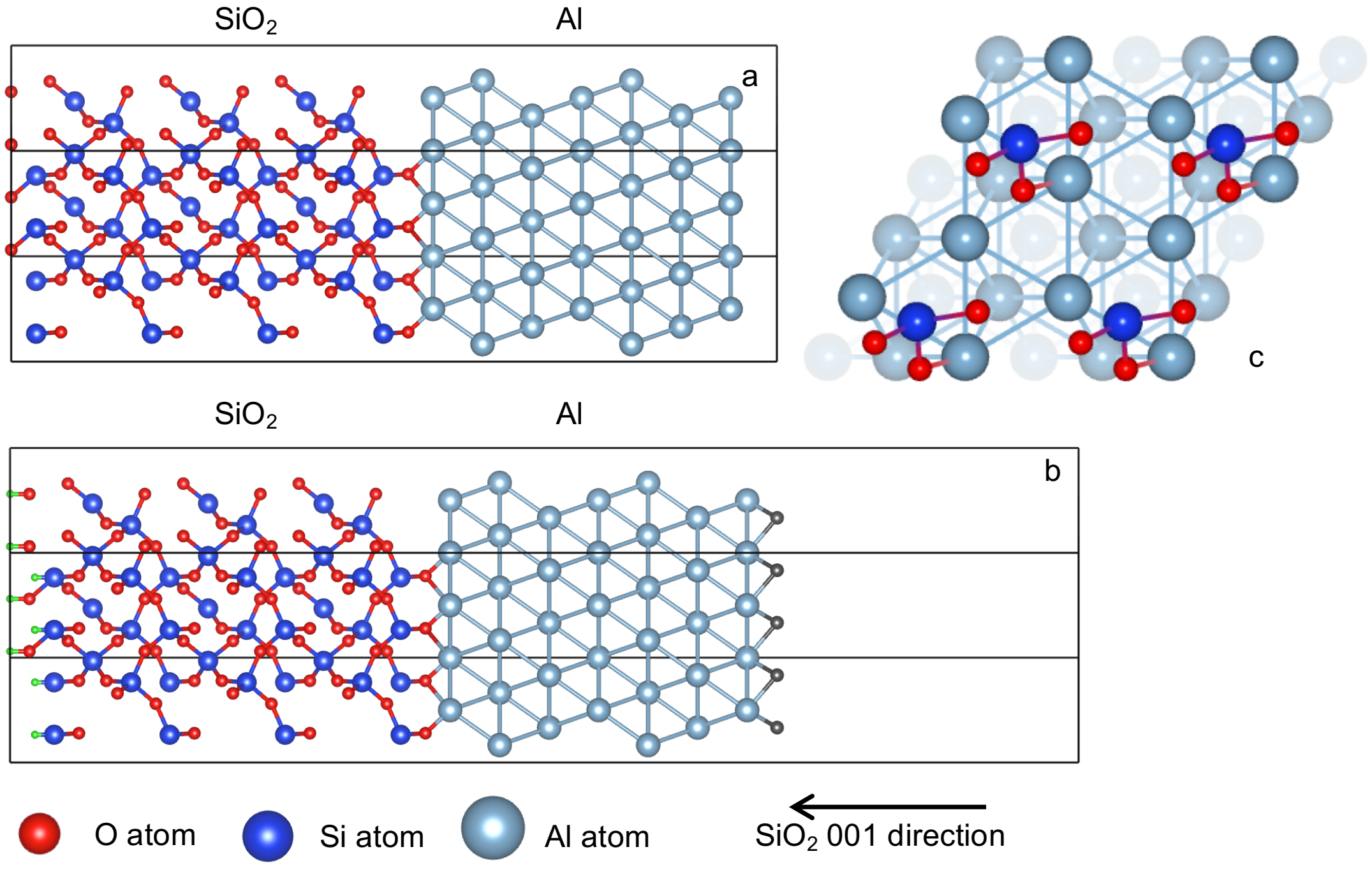

Figure 1 (color online): Sketches of the two types of supercells. Blue, red and grey balls show Si, O and Al atoms, respectively. a) Superlattice model ~10Å × 10Å × 35Å supercell containing 204 atoms. b) Slab model ~10Å × 10Å × 50Å supercell containing 224 atoms and a 15Å thick vacuum region. Surface passivating atoms H and 0.75 fractional H are shown by green and dark grey balls, respectively. c) Top view of the Al/$SiO_2$ interface showing the first 3 atomic layers around the interface.

To study the effects of metal-oxide interfaces on the materials, the charge density differences have been computed. The results reveal how charge densities in Al or $SiO_2$ alone are altered due to the presence of an interface. Three separate calculations are performed to obtain the charge density difference: (i) for the whole interface system $\rho_{Al/SiO_2}$, (ii) the system with only the Al part $\rho_{Al}$, and (iii) the system with only the $SiO_2$ part $\rho_{SiO_2}$. To obtain the charge density difference the $\rho_{Al}$ and $\rho_{SiO_2}$ parts are subtracted from the interface system charge density $\rho_{diff} = \rho_{Al/SiO_2} - \rho_{Al} - \rho_{SiO_2}$. Hence, $\rho_{diff}$ illustrates the charge density variations in Al and $SiO_2$ when they are put in contact.

Band alignments have been calculated using the Van de Walle and Martin method. [5] In this method, the macroscopic electrostatic potential is used as a reference for the electronic band structure. To calculate a band alignment, three different calculations are required: (i) a bulk Al calculation giving the metal Fermi level $E_F$ with respect to the bulk Al electrostatic potential, (ii) a bulk $SiO_2$ calculation giving the oxide conduction band minimum $E_{CBM}$ with respect to the bulk $SiO_2$ electrostatic potential, and (iii) an interface calculation giving the electrostatic potential offset $\Delta V$ between $SiO_2$ and Al when in contact. The band alignment of Al/$SiO_2$ is then constructed based on the different conduction levels using bulk macroscopic electrostatic potentials as references, and the potential offset at the interface. From the band line up, the electron potential barrier $\Phi = E_{CBM} - E_F$ is calculated. [6]

## 3) Results and discussion

In order to obtain quantitative results about the electronic properties of the Al/$SiO_2$ interface, the material models and computational setup need to be carefully tailored. Indeed, implicit periodic boundary conditions in plane wave DFT calculation codes needs to be accounted for when designing the supercells for the metal oxide interface study. Moreover, the band gap underestimation of the LDA also needs to be addressed to provide accurate electron potential barriers. Therefore the calculations for Al/$SiO_2$ interfaces have been performed using two different types of supercell. These two models are discussed in section 3a. Results obtained using conventional LDA and hybrid PBE0 exchange and correlation functionals are also discussed in section 3a. This ensures that the most reliable computational setup is chosen for the band line up study of defective interfaces in section 3b. Oxygen vacancies and their charge state at the metal-oxide interface are discussed in section 3b, along with their effect on electron potential barrier height.

**a. Band line up at defect free interface**

**i- Supercell models**

Band alignments at defect free Al/$SiO_2$ interface have been computed and are shown on Figure 2. The calculated barrier heights using LDA exchange and correlation functional are 3.43 and 2.78 eV for the superlattice and slab supercell models respectively. Following the Van de Walle and Martin method, the only difference between these two barrier heights is the electrostatic potential offset $\Delta V$. In order to investigate the discrepancy between the superlattice and slab model potential offsets, charge density differences have been computed for both supercell types, and are shown in Figure 3. For both supercell types, connecting the metal and the oxide parts leads to a charge density alteration that penetrates approximately 3-4 atomic layers deep inside the Al part. Therefore, in the superlattice supercell model a strong interface-interface coulombic interaction appears because the Al part is only 7 layers wide with two identical interfaces on both sides. In the slab supercell model however, the distance between the interface and its periodic images is larger because (i) it contains only one interface, and (ii) it contains an additional vacuum region. The use of a separating vacuum region effectively removes the spurious interface-interface interaction because, even though the Al part is still 7 layers wide, its bulk-like region is now wider. In this bulk-like region, the charge density difference is nil. Hence, slab results are deemed more reliable than the superlattice ones. In addition, sticking to the superlattice model but with thicker Al and $SiO_2$ regions to remove the spurious interaction is much more computationally demanding. It is worth noting that the LDA-superlattice calculated potential barrier height shows a good agreement with experimental measurements. [20,37-38] Since LDA tend to underestimate band gaps, this agreement is most likely fortuitous. Indeed, the calculated LDA band gap of bulk $SiO_2$ is 5.83 eV, which is 3 eV below its 8.9 eV experimental value. [8] The error on the LDA band gap is as large as the calculated barrier height, raising reliability issues.

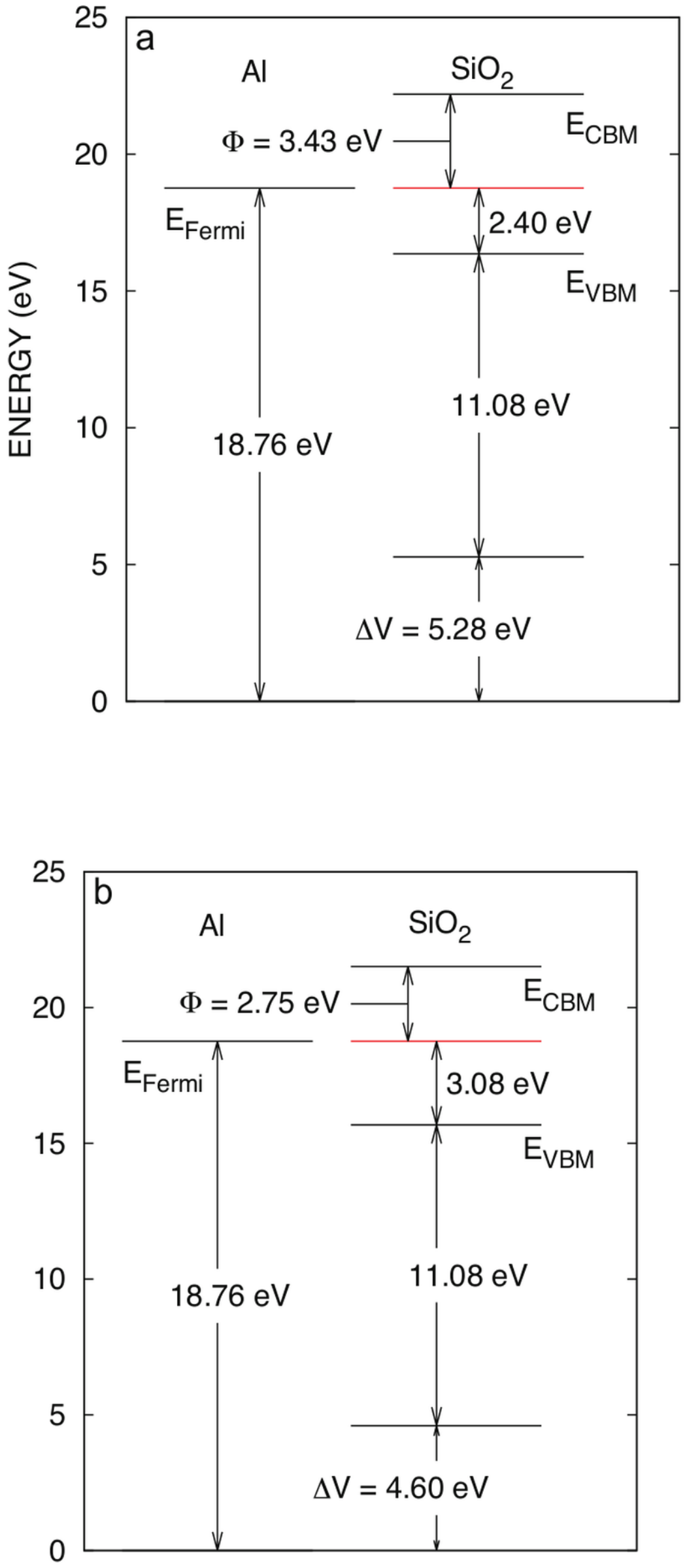

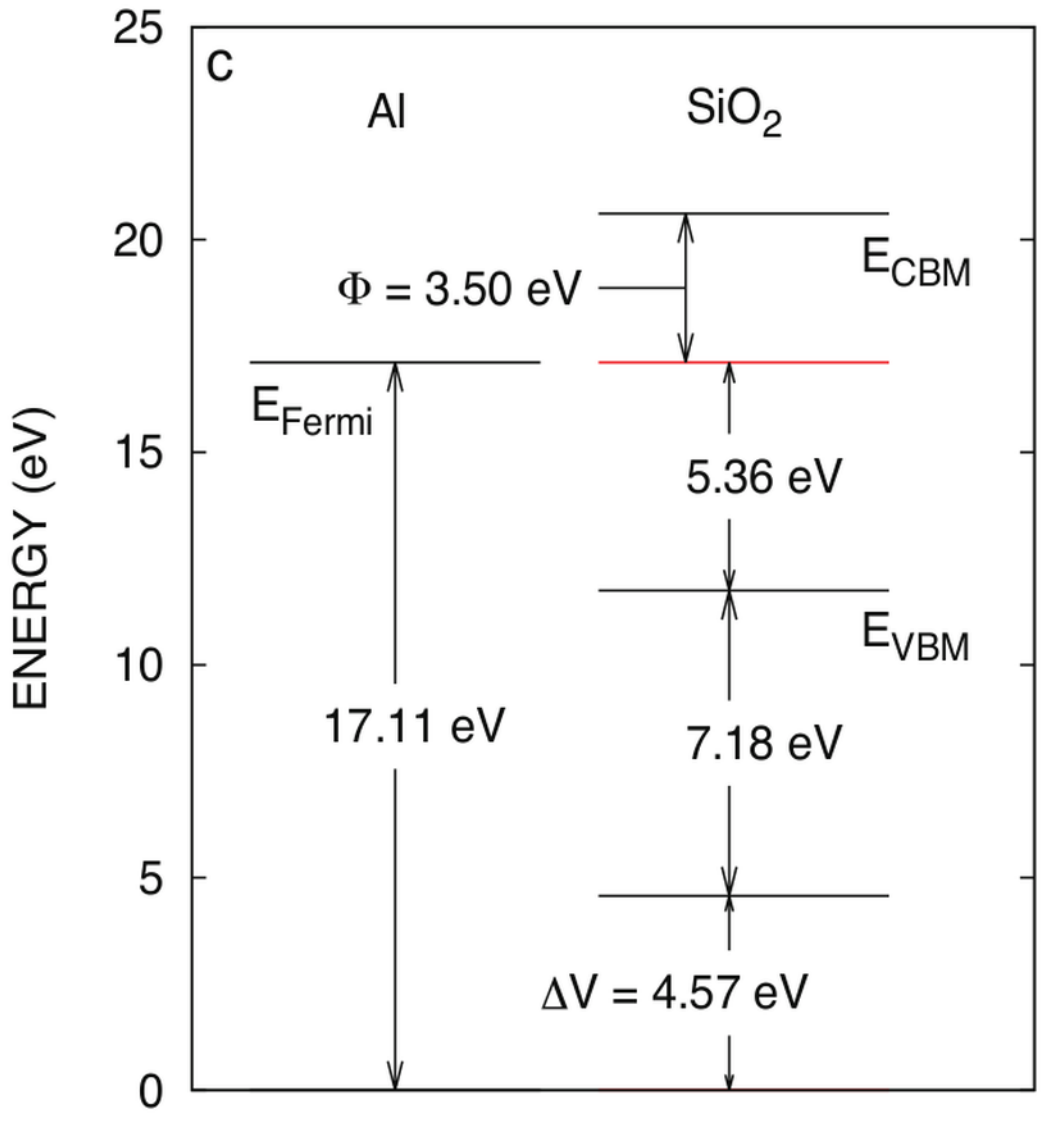

Figure 2: Band line up for defect free Al/$SiO_2$ interfaces, calculated using a) LDA exchange and correlation with superlattice model, b) LDA exchange and correlation with slab model, and c) PBE0 exchange and correlation with slab model.

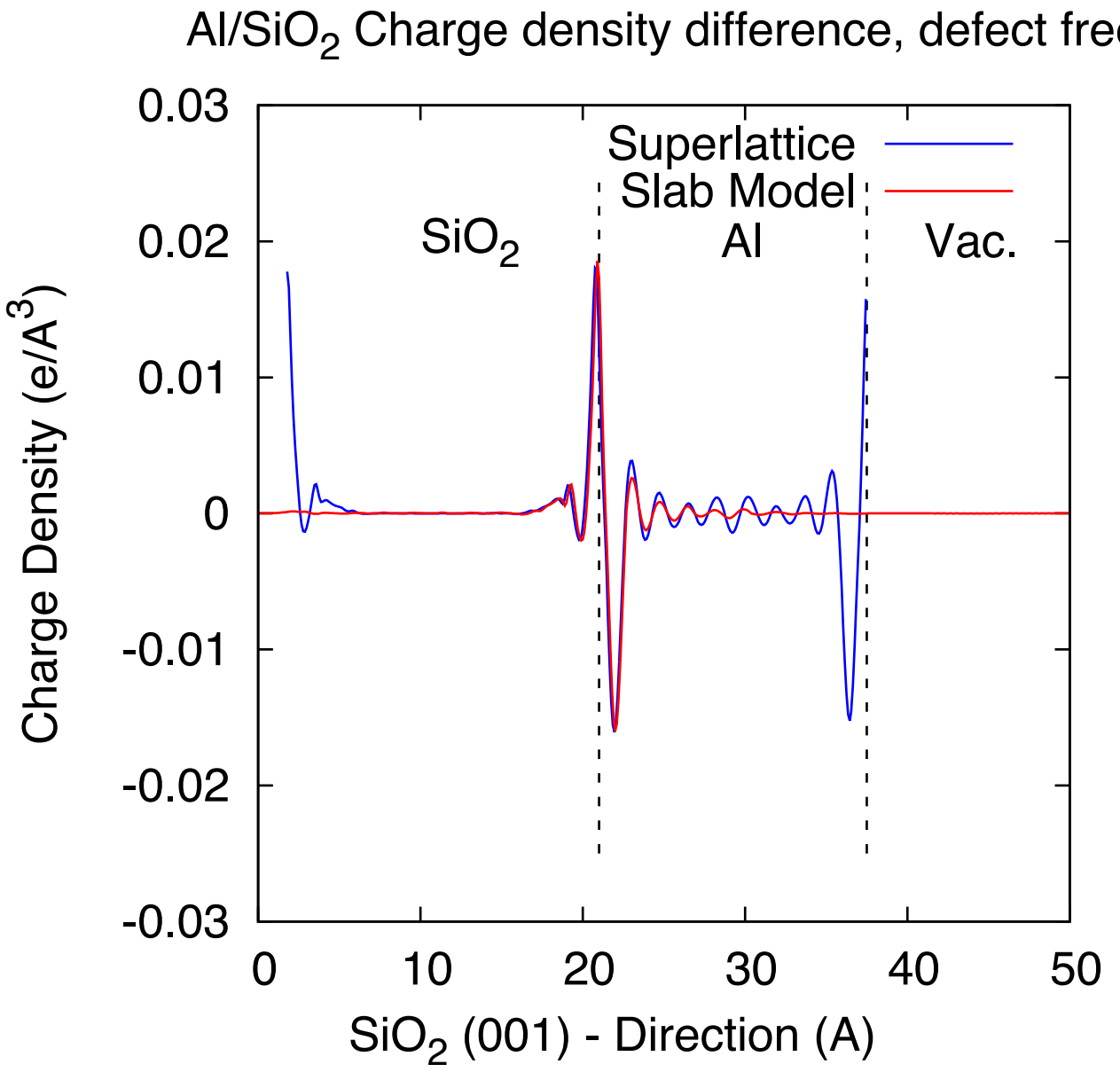

Figure 3 (color online): Charge density difference averaged in planes parallel to the interface, as a function of the plane position. Results for defect free Al/$SiO_2$ interfaces for superlattice (blue solid line) and slab supercell (red solid line, extending in the vacuum region) models. Vertical dashed lines separate different regions.

### ii- Exchange and correlation functional

The band alignment at a defect free Al/$SiO_2$ interface has been computed for the slab model supercell using the hybrid PBE0 exchange and correlation functional [12-13], and is shown on Figure 2c. All band alignment calculations have been done with PBE0: interface potential offset, bulk Al Fermi level, and bulk $SiO_2$ electronic structure. In $SiO_2$, the valence bands that have shifted downwards compared to LDA, have absorbed most of the 3 eV band gap correction. The calculated PBE0 $SiO_2$ band gap is 8.86 eV, in excellent agreement with experimental measurements giving 8.9 eV. [8] Indeed, admixing 25% of exact exchange in PBE0 partially correct for LDA self-interaction for occupied states, and hence lowers the valence band width. [29] Accordingly, the Fermi level in the Al part has also shifted downward. The calculated PBE0 and LDA electrostatic potential offset, 4.57 and 4.60 eV respectively, are very similar. This suggests that unlike band gaps, charge densities and bonding are already well described at the LDA level.

The calculated electron potential barrier height is 3.50 eV, in excellent agreement with experimental values given in the 3.4-3.6 eV range. [20,37-38] Since the slab supercell model does not suffer from the superlattice supercell model finite size effects, such as strong interface-interface interaction, this result supports that the agreement between experiments and the LDA superlattice model calculated potential barrier is fortuitous. It arises from error compensation, between finite size effects and band gap underestimation. Therefore, all following calculations are performed in slab type supercells with PBE0 exchange and correlation functional.

## b. Band line up at defective interface

### i- Charge distribution, and electrostatic potential offset

Oxygen vacancy ($V_o$) and oxygen di-vacancy ($V_{oo}$) have been investigated by removing one oxygen atom, and two oxygen atoms bonded to the same Si atom at the interface, respectively. The charge density difference caused by the presence of the oxygen vacancy has been calculated as $\rho_{diff} = \rho^{V_O}_{Al/SiO_2} - \rho^{defect-free}_{Al/SiO_2} - \rho_O$ , where $\rho^{V_O}_{Al/SiO_2}$ and $\rho^{defect-free}_{Al/SiO_2}$ are the charge densities at the defective and defect-free interface respectively, and $\rho_O$ is the charge density of a single oxygen atom. Analysis of the charge density difference in Figure 4a, shows a density increase between the Al and Si atoms through the oxygen vacancy, as if they were bonding. Further analysis of partial charge densities in Figure 4b, shows that unoccupied states are localized at the vacancy site. This suggests that oxygen vacancies could to trap free electrons as pointed out in previous studies. [9,10,17,22]

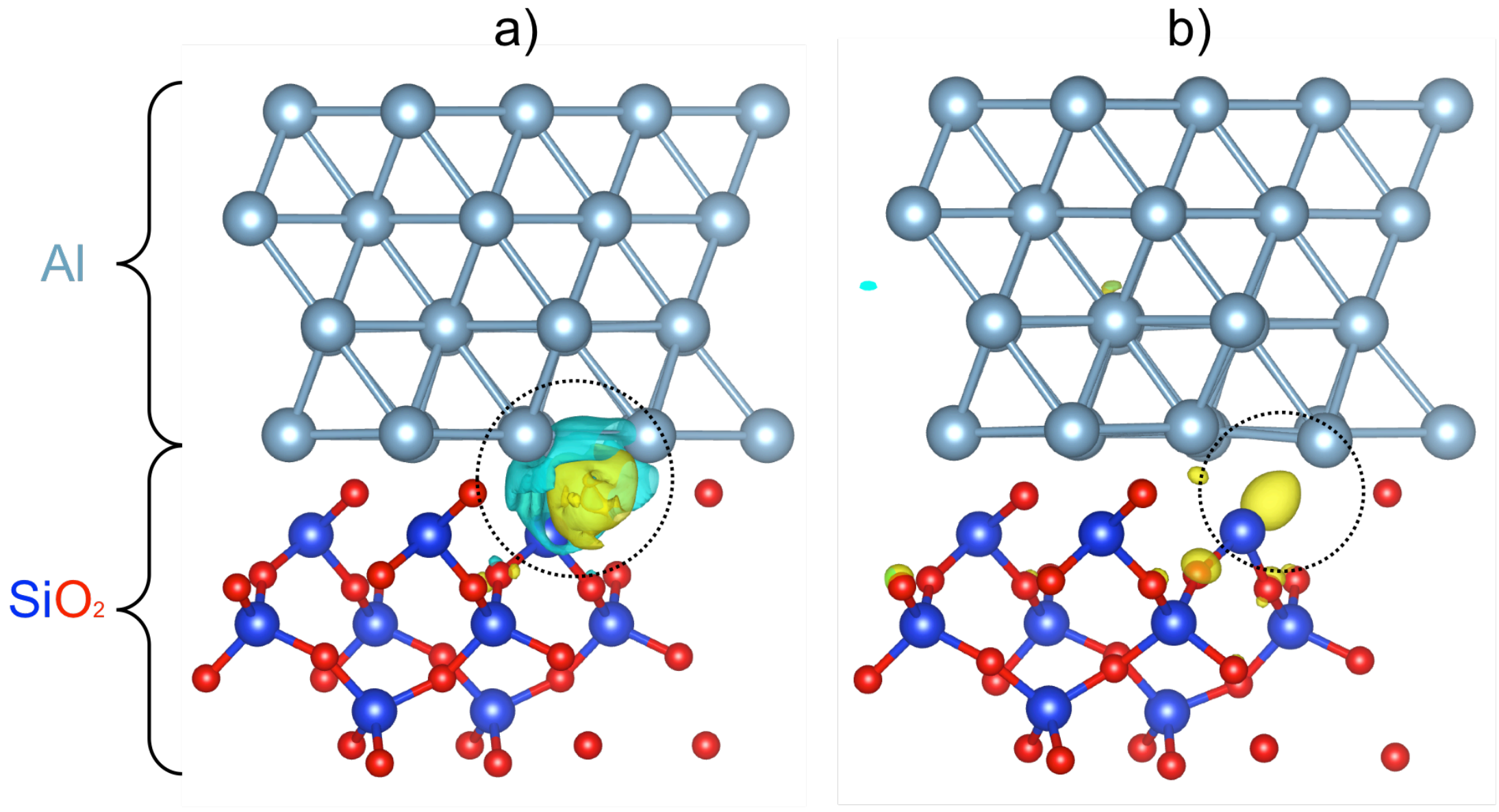

Figure 4 (color online): Defective interface. a) Charge density difference calculated as $\rho_{diff} = \rho^{V_O}_{Al/SiO_2} - \rho^{defect-free}_{Al/SiO_2} - \rho_O$, where blue and yellow regions correspond to charge density decrease and increase respectively. b) Charge density isosurface of a conduction state showing localization at the vacancy site.

The presence of a defect at the interface and the redistribution of charge surrounding it have deep implications on the determination of potential barrier height. In the slab supercell model, each Al and $SiO_2$ sub-slab has one interface and one surface, and are therefore asymmetric. The point defect creates an uneven distribution of charges at the sub-slabs endpoints, ultimately creating dipoles. This is shown on Figure 5a where a charge density peak is located at the oxygen vacancy site for both positively and negatively charged systems. Electric fields generated by dipoles bend averaged electrostatic potentials (see Figure 5b), otherwise flat throughout the supercell. It is worth noting that the average electrostatic potential in Al stays flat. Indeed, metals have mobile charges that screen dipoles. The electrostatic potential bending makes the determination of electrostatic potential offset not straightforward because the position of the interface is not known exactly between the two materials. [30] However, in the present study, dipoles are small compared to electrostatic potential offsets. Hence, we are not in a pathological case where the metal Fermi level artificially end up above the insulator conduction band minimum. [31] We assumed that the interface position lies midway between the metal and oxide atomic planes for simplicity. Indeed, we are looking at the offset between two straight lines fitted to the averaged electrostatic potentials in the bulk-like parts of $SiO_2$ and Al. Therefore, the determined electrostatic offset depends on the abscissa at which it is evaluated, that is the assumed position of the interface. By varying this assumed interface position by an amount given by half the interatomic plane distance, we estimate a maximum error of ±0.1 eV on the electrostatic potential offset determination, and hence on the potential barrier heights. Potential corrections as implemented in VASP [23-24,39] yields similar electrostatic potential offsets.

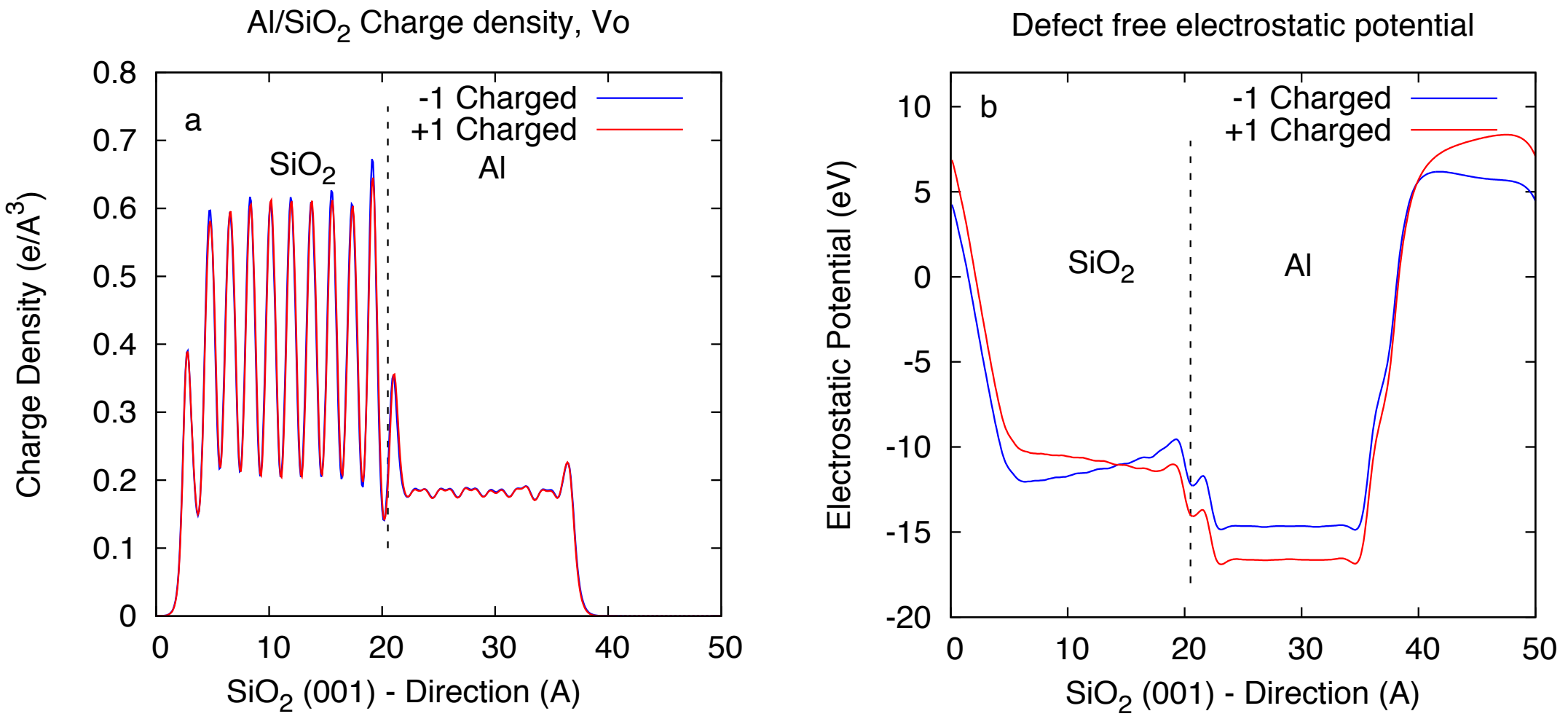

Figure 5 (color online): a) Valence charge density throughout a defective slab supercell. Positively (red) and negatively (blue) charged defects clearly show a charge density peak at the oxygen vacancy location. b) Average macroscopic electrostatic potential throughout a slab supercell for a positively (red) and negatively (blue) charged defect.

**ii- Defect formation energies**

Formation energies for slab interface supercells are shown on Figure 6. While not intended to, an overall agreement with experimental measurements of defect formation energies for oxygen vacancy and di-vacancy in bulk $SiO_2$ is found. [14,18,19] Analysis of the formation energies reveals the effect of Al on the interface defect: all $\varepsilon(+q_i/-q_i)$ charge state transitions, where the formation energies $\varepsilon(+q_i)$ and $\varepsilon(-q_i)$ are equal, occur at the same Fermi level for $q_i = 1,2,3$. Using the formation energy expression in Equation 2 for charge states $+q_i$ and $-q_i$ reveals that the total energy difference $E_{defect}^{+q_i} - E_{defect}^{-q_i}$ is equal to the total charge difference $2q_i$ times the electron chemical potential $\mu_e$. In other words, the energy cost of removing or adding an electron is equal to a constant reservoir chemical potential, irrespective of the initial charge state. The defect being located at the interface with Al, the metal certainly act as an electron reservoir for the defect. This analysis holds for other charge state transitions, e.g. $\varepsilon(+q_i/-q_j)$ with $q_i = 1,2,3$ and $q_j = -q_i + 1$.

Our band line up results using Van de Walle method (Figure 2c) show that Al Fermi level is set at 5.36 eV above the defect free $SiO_2$ VBM (Valence Band Maximum). In Van de Walle method, the aluminum Fermi level and $SiO_2$ VBM are both bulk properties. The only purpose of the interface supercell calculation is to determine the electrostatic potential offset across the interface. No other interface information is provided. Directly analyzing the interface supercell projected density of states confirms that the supercell Fermi level is an aluminum state, not an interface state, and is set at 5.3 eV above the $SiO_2$ VBM. These states are identified as the highest occupied states that projects onto Al and $SiO_2$ atoms respectively. Therefore, the interface supercell VBM, which is used as the energy reference and set at 0 eV in Figure 6, is the aluminum Fermi level. This Fermi level is well above the $SiO_2$ mid-gap energy. For a Fermi level at 1 eV above the supercell VBM, or equivalently 6.3 eV above the $SiO_2$ VBM, negatively charged oxygen vacancies are more likely. It is worth noting that around the aluminum Fermi level energy (0 eV in Figure 6), oxygen vacancies are likely to be neutral. However, increasing or decreasing the Fermi level will make negatively or positively charged oxygen vacancies more likely, respectively. This suggests that oxygen vacancies can act both as donors or acceptors depending on the Fermi level. Oxygen di-vacancy formation energies are roughly twice those of single oxygen vacancy. Therefore, oxygen di-vacancies occurrence at the Al/$SiO_2$ interface is rare.

Oxygen vacancies are best known for their hole trapping abilities in bulk $SiO_2$, not electron trapping. However, the current study focuses on defects at the interface between $SiO_2$ and a metal. In this case, the proximity of the metallic contact infers the control of the vacancy charge state to the metal via the control of the Fermi level. This gives the oxygen vacancy the possibility to be extrinsically negatively charged. Moreover, a recent study showed that negatively charged oxygen vacancies in $SiO_2$ can explain enhanced electron trapping observed in SiC MOSFETs [41]. The defect charging process is described and linked to actual functioning temperature of the device. Also, their equivalent areal oxygen vacancy density used for their calculations is the same as ours.

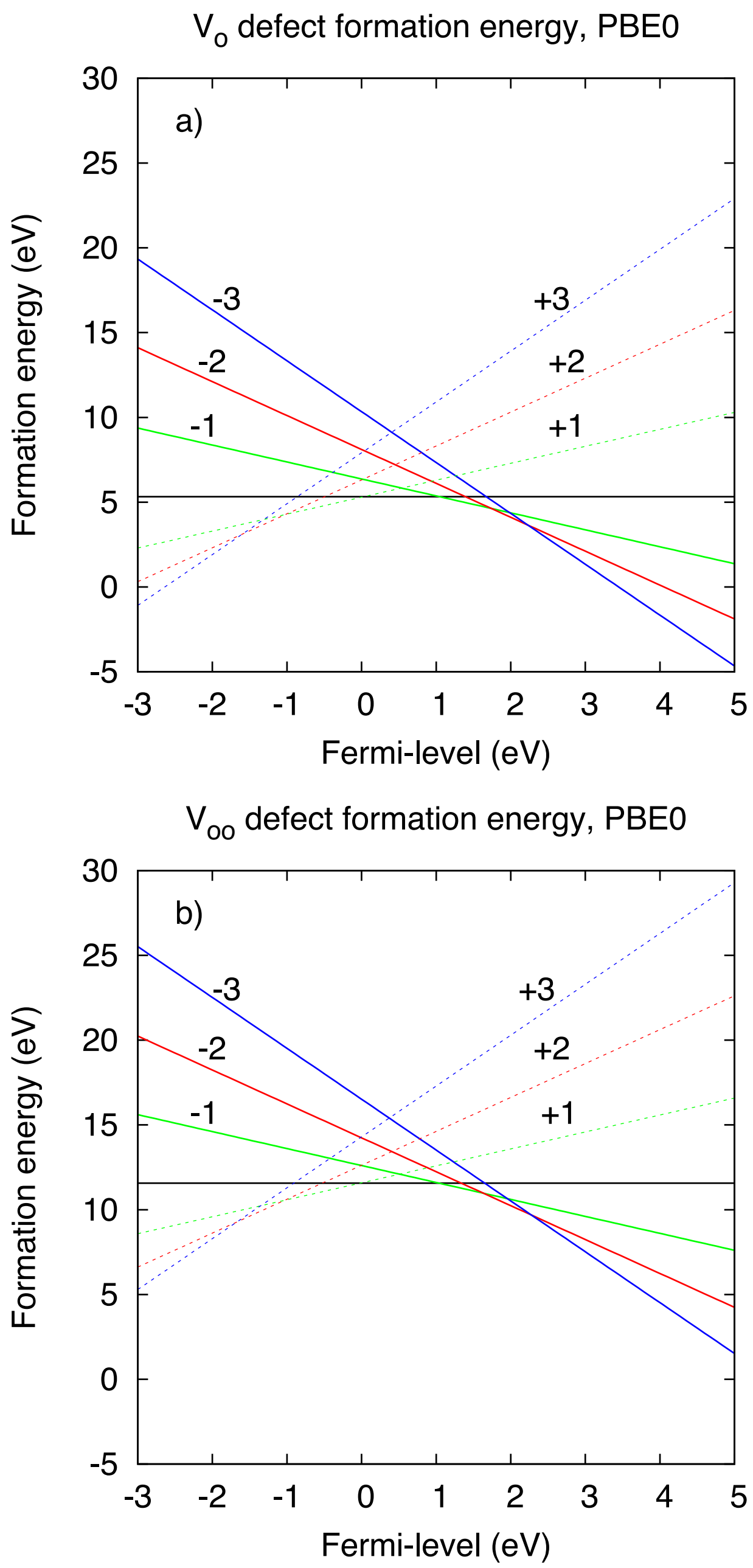

Figure 6: PBE0 Interface defect formation energies: (a) oxygen vacancy; (b) oxygen di-vacancy. Horizontal black solid lines refer to neutral charge states. The zero Fermi level has been set at the neutral oxygen vacancy supercell VBM energy that corresponds to Al Fermi level.

**iii- Band line up with charged defects**

Table I summarizes the calculated potential barrier height for a defective interface as a function of the defect charge state. It is found that the potential barrier height varies with the defect charge state. Indeed, more electronically charged defects leads to stronger repulsive Hartree interaction. Since the interfacial oxygen vacancies are linked to the metal Al layer, as suggested by the charge densities and defect formation energy studies, the stronger Hartree interaction pushes the average electrostatic potential upwards at the interface and in the metal Al layer. This results in a decrease of the electrostatic potential offset $\Delta V$ between Al and $SiO_2$, and therefore leads to a decrease of the electron potential barrier $\Phi$. In the oxygen vacancy case, the electron potential barrier height drops from 3.68 eV to 3.34 eV when the vacancy charge state goes from neutral to -1. Negatively charged oxygen vacancies at the interface results in lower electron potential barrier heights compared to the defect free interface (3.50 eV).

Just like oxygen vacancies, oxygen di-vacancies also strongly alter the band line up at $SiO_2$/Al interfaces, and follow the same charge state trend. Recalling Equation 1, smaller electron potential barrier heights would increase the barrier tunnel transmission coefficient and hence contribute to larger tunneling currents. Compared to defect free interfaces, smaller electron potential barrier heights occurs for negatively charged oxygen vacancies. Since oxygen vacancies can trap conduction electrons, they constitute weak spots at the interface: tunneling electrons can be trapped at the vacancy site and lower the barrier potential height, leading to even more tunneling electrons until breakdown occurs.

It is worth noting that calculated electron potential barrier height decreases when increasing the oxygen vacancy concentration. Going from a defect free case, to the maximum concentration decreases the barrier height by 0.2 eV. This is the maximum range, into which the barrier height varies by varying the defect concentration. Our current defect concentration value of $1.2\times10^{14}cm^{-2}$, sets our typical defective interface material picture in the middle of this widow. If our barrier height results were to be interpreted in the dilute limit and not as $1.2\times10^{14}cm^{-2}$ interfacial oxygen vacancy concentrations, we expect the barrier heights to be off by roughly 0.1 eV. Hence, if the defect concentration were to be totally disregarded, this would translate into an extra 0.1 eV to our current error window on the absolute value of electron potential barrier heights.

| Unit in eV. | Defect Free | $V^{+3}$ | $V^{+2}$ | $V^{+1}$ | $V^{neutral}$ | $V^{-1}$ | $V^{-2}$ | $V^{-3}$ |
|---|---|---|---|---|---|---|---|---|
| $\Phi_{Vo}$ | 3.50 | 4.37 | 4.11 | 3.92 | 3.68 | 3.34 | 3.13 | 2.90 |
| $\Phi_{Voo}$ | | 4.49 | 4.15 | 3.98 | 3.73 | 3.47 | 3.30 | 3.08 |

Table I: Calculated potential barrier height with interfacial oxygen vacancy and di-vacancy defects. Defect charge states are indicated in superscripts. Calculations performed with PBE0 exchange and correlation functional and the slab interface supercell model.

**Conclusion**

By analyzing the Al/$SiO_2$ interface with state of art DFT calculations, the effect of oxygen vacancies at the metal-oxide interface on electron potential barrier height has been assessed. A strong dependence of the barrier height on the oxygen vacancy charge state has been found. More specifically, it is found that oxygen vacancies or oxygen di-vacancies in negative charge states leads to lower electron potential barrier heights compared to a defect-free interface. Our charge densities and defect formations energy analysis support that oxygen vacancies at the Al/$SiO_2$ interface can trap electrons and be in negative charge states. In the rare event scenario, negatively charged oxygen vacancies or di-vacancies, which lower electron potential barrier heights, can lead to larger tunneling current through the dielectric that can furthermore lower barriers height if electrons are trapped at the vacancy site. This destructive feedback loop can lead to excessive tunneling current through the dielectric until breakdown occurs.

**Acknowledgement**

This work was funded by the Air Force with program name: Aerospace Materials for Extreme Environment, and grant number: FA9550-14-1-0157. We acknowledge Advanced Research Computing at Virginia Tech for providing computational resources and technical support that have contributed to this work (http://www.arc.vt.edu).